# Synthesis of Component and Connector Models from Crosscutting Structural Views

Shahar Maoz
School of Computer Science
Tel Aviv University, Israel

Jan Oliver Ringert, Bernhard Rumpe
Software Engineering
RWTH Aachen University, Germany

## ABSTRACT

We present component and connector (C&C) views, which specify structural properties of component and connector models in an expressive and intuitive way. C&C views provide means to abstract away direct hierarchy, direct connectivity, port names and types, and thus can crosscut the traditional boundaries of the implementation-oriented hierarchical decomposition of systems and sub-systems, and reflect the partial knowledge available to different stakeholders involved in a system's design.

As a primary application for C&C views we investigate the synthesis problem: given a C&C views specification, consisting of mandatory, alternative, and negative views, construct a concrete satisfying C&C model, if one exists. We show that the problem is NP-hard and solve it, in a bounded scope, using a reduction to SAT, via Alloy. We further extend the basic problem with support for library components, specification patterns, and architectural styles. The result of synthesis can be used for further exploration, simulation, and refinement of the C&C model or, as the complete, final model itself, for direct code generation.

A prototype tool and an evaluation over four example systems with multiple specifications show promising results and suggest interesting future research directions towards a comprehensive development environment for the structure of component and connector designs.

## Categories and Subject Descriptors

D.2.1 [**Software Engineering**]: Requirements/Specifications; D.2.2 [**Software Engineering**]: Design Tools and Techniques

## 1. INTRODUCTION

Component and connector (C&C) models are used in many application domains, from cyber-physical and embedded systems to web services to enterprise applications. The structure of a C&C model consists of components at different containment levels, their typed input and output ports, and the connectors between them.

A system's C&C model is typically complex; it is not designed by a single engineer and is not completely described in a single document. Moreover, some C&C models may be bound to reuse library or third-party components designed and documented elsewhere. Thus, we consider a setup where many different, incomplete, relatively small fragments of the model are provided by architects responsible for subsystems, for the implementation of specific features, use cases, or functionality, which crosscut the boundaries of components. Such fragments may be developed by separate, distributed teams, each focusing on only some aspects of the system and its development and having only partial knowledge of the system as a whole. Moreover, a team may have several, alternative solutions that address the same concern, and some knowledge about designs that must not be used. To move forward in the development process and enable implementation, these partial models and the intentions behind them should be integrated and then realized into a single, complete design. However, such an integration is a complex and challenging task.

In this paper we present *component and connector views*, which specify structural properties of component and connector models in an expressive and intuitive way. C&C views provide means to abstract away direct hierarchy, direct connectivity, port names and types. Specifically, C&C views may not contain all components and connectors (and typically indeed contain only a small subset of the set of all components and connectors of the system, related only to a specific use case or set of functions or features). They may contain (abstract) connectors between components at different, non-consecutive containment levels, and they may provide incomplete typing information, that is, components' ports may be un-typed. While the standard structural abstraction and specification mechanisms supported by existing languages and tools for C&C models rely on the traditional, implementation-oriented hierarchical decomposition of systems to sub-systems, we have defined C&C views to allow one to specify properties that crosscut the boundaries of sub-systems. Most importantly, this makes them especially suitable to reflect the partial knowledge available to different stakeholders involved in a system's design.

As a primary application for C&C views we investigate the synthesis problem: given a C&C views specification, consisting of mandatory, alternative, and negative views, construct a concrete satisfying C&C model, if one exists.

We show that the synthesis problem for C&C views specifications is NP-hard and solve it, in a bounded scope, using a reduction to SAT, via Alloy [20] (since the problem is NP-hard, the use of Alloy/SAT to solve it is justified). The generated Alloy module is translated to a Boolean formula and solved by a SAT solver. If a satisfying assignment is found, we translate it back to a complete C&C model and present it to the engineer; this design is ready for code generation. All analysis is fully automated and is internally done using Alloy's APIs and embedded SAT solver. The engineer need not see the generated Alloy module.

The input for the synthesis is a C&C views specification. Its output is a single C&C model that satisfies the specification and is complete, to allow implementation. When more than one solution exists, the engineer can explore different alternatives. When no solution exists (within a bounded scope), the technique reports that the input specification is unsatisfiable.

As a concrete language for C&C models we use MontiArc [2,16], a textual ADL developed using MontiCore [23], with support for direct Java code generation (including interfaces, factories, etc.). Its expressive power is comparable to that of other ADLs, e.g., MathWorks Simulink [26], AADL [9], and UML Component diagrams. The C&C views are defined as an extension to general C&C models. The concrete syntax used in our implementation is an extension of MontiArc.

To further increase the usefulness of C&C views synthesis in practice, we have extended the basic synthesis problem with support for three advanced features. First, support for integration with pre-defined or library components. Second, support for several high-level specification patterns. Third, support for synthesis subject to several architectural styles. We report on these advanced features in Sect. 5.

We have implemented C&C views synthesis, including the more advanced extensions mentioned above, and evaluated it by applying it to four example systems. We report on our implementation and evaluation in Sect. 6. The implementation and example specifications reported on are available in supporting materials [3].

Some previous work deal with the analysis and synthesis of component and connector structures, mainly in the context of architectures [6,7,13,19,21]. Other work deal with synthesizing behavior rather than structure [17,24,31,32]. We discuss these and other related work in Sect. 7.

Sect. 2 gives a semi-formal overview of C&C views and synthesis using examples. Sect. 3 provides formal definitions and Sect. 4 describes our solution. Advanced support for integration of library components, specification patterns, and several architectural styles, is described in Sect. 5. The implementation and evaluation are presented in Sect. 6. Sect. 7 discusses related work. Sect. 8 concludes.

## 2. OVERVIEW

We present an overview of C&C views synthesis using an example, adopted from an industrial model of a robot arm[1], which is typical to a cyber-physical or an embedded system.

---
[1] We thank Ali Muhammad, Remote Operation and Virtual Reality Group, VTT Tampere, Finland, for allowing us to use this model.

We focus here on a single joint of this arm. The example is discussed semi-formally. Formal definitions are given in Sect. 3.

### 2.1 Example I

Fig. 1 shows the C&C views specification $S_1$, consisting of six views. The C&C view `RJFunction` describes the system architecture from the point of view of the team responsible for its function: the `RotationalJoint` contains a `Cylinder` and a `Sensor` that is connected to an `Actuator`. As a C&C view, `RJFunction` is partial, so it may not contain all the system's components. Moreover, while the components shown inside the joint must actually be inside the joint, they may be nested within some of its subcomponents (not shown in this model). On the other hand, the C&C view specifies that the three subcomponents, `Cylinder`, `Sensor`, and `Actuator` are not nested within one another. Finally, `Sensor` and `Actuator` must be connected, but their connection is not necessarily direct and the port names and types are not given in the view.

The C&C views `BodySensorIn` and `BodySensorOut` describe two alternatives for the C&C model from the point of view of the team responsible for a component named `Body` and focus on its internal structure. The first specifies four subcomponents of `Body` (not necessarily direct sub components, not necessarily all of them) and the connections between them (again, not necessarily all connections, not necessarily direct ones). The second suggests an alternative, where `Sensor` is outside `Body`.

The C&C view `SensorConnections` describes the point of view of the engineer responsible for `Sensor`. It shows that `Sensor` is connected to `Cylinder` and to a component named `JointLimiter`. Again, the C&C view is partial, thus in the complete model the connections shown may be indirect and `Sensor` may be connected to additional components.

The C&C view `RJStructure` provides a high-level description of the `RotationalJoint` structure, some of the components it contains and the connections between them. It describes the knowledge of the senior engineer responsible for the joint. It also shows the name `angle` and type `float` of an incoming port of the `Cylinder` for a connection (not necessarily direct) coming from `Body`.

Finally, the C&C view `ASDependence` shows the `Actuator` and the `Sensor` inside `Body`. It describes some domain knowledge concerning a requirement for independence between `Actuator` and `Sensor`. Thus, it is used in the specification (see below) in a negated form, to not allow an architecture where `Actuator` and `Sensor` are both inside the same component, in our case, `Body`.

The Boolean expression for the C&C views specification $S_1$ is `RJFunction` $\land$ (`BodySensorIn` $\lor$ `BodySensorOut`) $\land$ `SensorConnections` $\land$ $\lnot$`ASDependence` $\land$ `RJStructure`. Is there a complete C&C model that satisfies this specification? Our work provides a fully automated and constructive answer to this question. Specifically, given this specification, our tool provides a positive answer and outputs the complete C&C model shown in Fig. 2. For readability, we omit some port names, types, and internal connectors from the diagram.

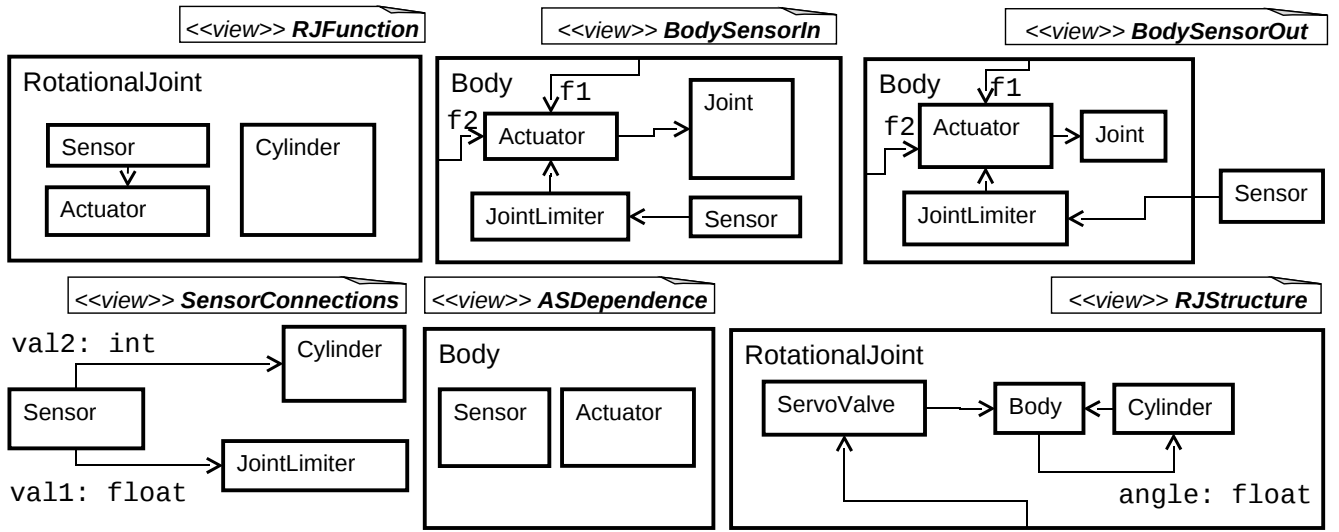

**RJFunction AND (BodySensorIn OR BodySensorOut) AND SensorConnections AND NOT ASDependence AND RJStructure**

Figure 1: A C&C views specification $S_1$. Note that the specification states that BodySensorIn and BodySensorOut are alternatives (at least one needs to be satisfied) and that ASDependence is negated, so it must not be satisfied by the model.

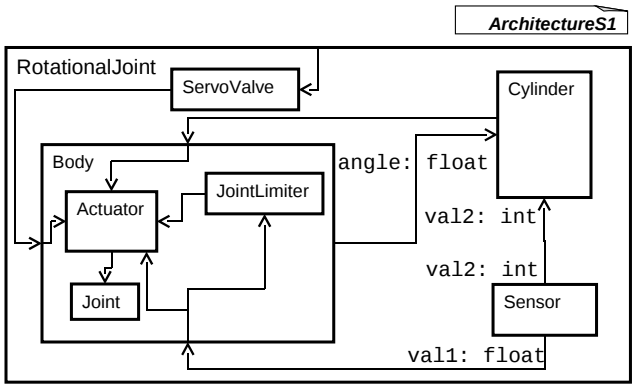

Figure 2: A C&C model with 19 ports satisfying the C&C view specification $S_1$.

## 2.2 Example II

Consider now the additional C&C view OldDesign and revised specification $S_2$ shown in Fig. 3. The view specifies that Actuator is connected to Cylinder and that both components are contained inside Body (although not necessarily directly). It also shows the name angle and type int of the Cylinder's incoming port for a connection (not necessarily direct) coming from Actuator.

Is there a complete C&C model that satisfies the revised specification $S_2$ (consisting of $S_1$ after adding OldDesign as another conjunct)?

Our tool identifies that $S_2$ is unsatisfiable and informs that a complete C&C model that satisfies it does not exist. One reason relates to the containment relation between Body and Cylinder: according to the Structure C&C view, the two components are not contained within one another; according to the OldDesign C&C view, the latter is contained within the former. Another reason is the type conflict float vs. int for the Cylinder's incoming port angle.

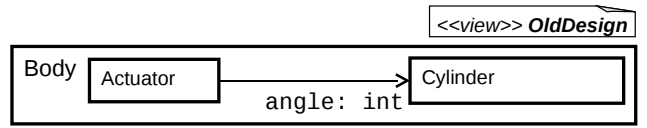

Figure 3: A C&C views specification $S_2$ extends specification $S_1$ (Fig. 1) with the additional view Old-Design shown here, added as another conjunct.

## 3. DEFINITIONS

We define the structure of C&C models and views as used in this paper (we give shortened definitions, for complete definitions see the technical report available from [3]).

### 3.1 Component and Connector Models

A C&C model is a structure
$cncm = \langle Cmps, Ports, Cons, Types, subs, ports, type \rangle$ where

- $Cmps$ is a set of named components, each of which has a set of ports $ports(cmp) \subseteq Ports$ and a (possibly empty) set of immediate subcomponents $subs(cmp) \subset Cmps$,
- $Ports$ is a disjoint union of input and output ports $Ports = PortsIn \cup PortsOut$ where each port $p \in Ports$ has a name, a type $type(p) \in Types$, and belongs to exactly one component $p \in ports(cmp)$,
- $Cons$ is a set of directed connectors, each of which connects two ports of the same type, which belong to two sibling components or to a parent component and one of its immediate subcomponents, and
- $Types$ is a finite set of type names.

Some additional well-formedness rules apply, e.g., that the subcomponents relation is a strict partial order, that every port has at most one incoming connector, and that port names are unique within their component. In addition, without loss of generality, we consider only C&C models with exactly one top component.

## 3.2 C&C views

A C&C view is a structure $view = \langle Cmps, Ports, AbsCons, Types, subs, ports, type \rangle$ where

- $Cmps$ is a set of named components, each of which has a (possibly empty) set of ports $ports(cmp) \subseteq Ports$ and a (possibly empty) set of subcomponents $subs(cmp) \subset Cmps$,
- $Ports$ is a disjoint union of sets of input and output ports $Ports = PortsIn \cup PortsOut$ where each port $p \in Ports$ has a (possibly unknown) name, a (possibly unknown) type $type(p) \in Types \cup \bot$, and belongs to exactly one component $p \in ports(cmp)$,
- $AbsCons$ is a set of abstract connectors, each of which connects components (optionally) via ports of the same type or an unknown type, and
- $Types$ is a finite set of type names.

Note that in a C&C view, abstract connectors are not required to connect only two sibling components or a parent component and one of its immediate subcomponents. Again, the subcomponents relation is a strict partial order, but we do not restrict C&C views to have exactly one top component. We are now ready to define the semantics of our C&C view and C&C model, and specifically, when does the second satisfy the first.

A C&C model satisfies a C&C view iff the types, components, and ports mentioned in the second are contained in the first, the first respects the subcomponent relation induced by the second, two ports connected by an abstract connector in the second are connected by a chain of connectors in the first (respecting direction, names, and types), and all ports of a component in the second belong to the same component in the first with corresponding name, type and direction. More formally:

DEFINITION 1 ($cncm \models view$). *A C&C model cncm satisfies an C&C view view iff:*

- $view.Types \subseteq cncm.Types$, $view.Cmps \subseteq cncm.Cmps$, $view.Ports \subseteq cncm.Ports$,
- $\forall cmp_1, cmp_2 \in view.Cmps: cmp_1 \in view.subs(cmp_2)$ iff $cmp_1 \in cncm.subs^+(cmp_2)$ (we use $^+$ to denote the transitive closure),
- $\forall ac \in view.AbsCons \exists$ chain of connectors in cncm, $c_1, \ldots, c_n$ with $ac.srcCmp = c_1.srcCmp$ and $ac.tgtCmp = c_n.tgtCmp$ with matching port names and types, if specified, and
- $\forall cmp \in view.Cmps$:
  (1) $view.ports(cmp) \subseteq cncm.ports(cmp)$, and
  (2) $\forall p \in view.ports(cmp): p \in view.PortsIn$ iff $p \in cncm.PortsIn \wedge view.type(p) \in \{\bot, cncm.type(p)\}$
  (similarly for unknown and given port names).

DEFINITION 2 (view SEMANTICS). *The semantics of a C&C view consists of the possibly infinite set of C&C models that satisfy it. Formally:*
$sem(view) = \{cncm \mid cncm \models view\}$.

A C&C views specification $S$ consists of a Boolean expression over a set of C&C views $V$. By simple extension, a C&C model $cncm$ satisfies a specification $S$, denoted $cncm \models S$, iff replacing each view $v$ in $S$ with the value of $cncm \models v$ makes $S$ true: $cncm \models S \Leftrightarrow S[v/(cncm \models v)]_{v \in V}$.

## 3.3 Problem Definition

The C&C views synthesis problem is defined as follows: given a C&C views specification $S$, find a C&C model $cncm$ s.t. $cncm \models S$ if such a model exists.

## 4. C&C VIEWS SYNTHESIS

### 4.1 Is C&C Views Synthesis Hard?

We show that the C&C views synthesis problem (even without connectors) is NP-hard, using a reduction from 3SAT. We give an overview below.

Given a 3SAT formula over variables $x_1, \ldots, x_n$ in clauses $c_1, \ldots, c_m$ we construct the following C&C views specification. First, for each variable $x_i$ we define two components $xT_i$ and $xF_i$ and two C&C views $vT_i$ and $vF_i$ such that in $vT_i$, $xF_i$ contains $xT_i$, and in $vF_i$, $xT_i$ contains $xF_i$. Intuitively, $vT_i$ ($vF_i$) represents a positive (negative) valuation for $x_i$. Obviously, a given C&C model can only satisfy one of them. Second, for each clause $c_j$ we create an C&C views clause $cV_j$ that includes a disjunction of three C&C views, $vT_i$ or $vF_i$ for each variable $x_i$ in $c_j$: if $x_i$ appears positive in $c_j$ we use $vT_i$, if it appears negative we use $vF_i$.

We create a Boolean expression for the C&C views specification consisting of a conjunction of (1) $xT_i \vee xF_i$ for all $1 \leq i \leq n$, and (2) C&C views clauses $cV_j$ for all $1 \leq j \leq m$. It is easy to see that the 3SAT formula has an assignment iff the C&C views specification has a satisfying C&C model.

### 4.2 Formulating the Problem in Alloy

Our solution is based on a reduction to Alloy. The Alloy module is analyzed using a SAT solver. If an assignment is found, we translate it back to a satisfying C&C model. The formulation of the C&C views synthesis problem in Alloy consists of four parts: (1) a fixed set of signatures and facts describing a meta-model for C&C models, (2) a fixed set of predicates used as a language to specify the semantics of C&C views, (3) a set of signatures and predicates derived from the specific input C&C views, and (4) the specification's Boolean expression.

List. 1 shows (part of) the Alloy code describing the meta-model for C&C models. For example, note the signatures `Component` and `Port`. A component has a set of ports, a set of sub components, and at most one parent. The fact `subComponentsAndParents` specifies the value for the field `parent`, and the fact `subComponentsAcyclic` specifies that the sub component relation is acyclic. Finally, the fact `portsOfComponentHaveUniqueNames` specifies that no two ports of the same component have the same name.

List. 2 shows (part of) the Alloy code used as a language to specify the semantics of C&C views. For example, the predicate `contains` is true if the child is in the transitive closure of the parent's sub component relation, and the predicate `independentSet` defines the semantics of a set of components where no two components contain each other. The predicate `connected` defines directed connectedness based on the transitive closure of the receiving port relation. The remain-

```
1  abstract sig Component {
2    ports : set Port,
3    subComponents : set Component,
4    parent : lone Component }
5
6  fact subComponentsAndParents {
7    all ch, par : Component|
8      (ch in par.subComponents iff ch.parent = par) }
9
10 fact subComponentsAcyclic {
11   no comp : Component |
12     comp in comp.^subComponents }
13
14 sig Port {
15   type : one Type,
16   name: one PortName,
17   direction: one Direction,
18   receivingPorts : set Port,
19   owner : one Component,
20   sendingPort : lone Port }
21
22 fact portsOfComponentHaveUniqueNames {
23   all c:Component | all disj p1, p2 : c.ports |
24     p1.name != p2.name }
```

**Listing 1:** Excerpts from the meta-model for C&C models in Alloy (see Sect. 4.2)

```
1  pred contains [parent: Component,
2                 child: Component] {
3    child in parent.^(subComponents) }
4
5  pred independentSet [components : set Component] {
6    all disj c1, c2 : components |
7      ( (no c1.subComponents)   or
8        (not contains[c1, c2]) ) }
9
10 pred connected[sender: Component,
11                receiver: Component] {
12   some p : receiver.ports |
13     p in sender.ports.^receivingPorts }
14
15 pred connectedWithPortNames[sender: Component,
16          sendName : PortName, receiver: Component,
17                      recvName: PortName] {
18   some sp : sender.ports |
19         some rp : receiver.ports |
20     rp.name = recvName and
21     rp in sp.^receivingPorts and
22     sp.name = sendName and
23     sp in rp.^~receivingPorts }
24
25 pred connectedWithReceiverPortName[
26          sender: Component, receiver: Component,
27                       recvName : PortName] {
28   some sendName : sender.ports.name |
29     connectedWithPortNames[sender, sendName,
30                       receiver, recvName] }
31
32 pred untypedPort[cmp : Component, dir: Direction,
33                       portName : PortName] {
34   some port : cmp.ports |
35     port.direction = dir and
36     port.name = portName }
```

**Listing 2:** Excerpts from the 'language' for C&C views' semantics (see Sect. 4.2)

ing predicates in List. 2 are used to express the semantics of abstract connectors and untyped ports.

For the third part, we collect from the input C&C views the component names, port names, and types into signatures extending `Component`, `PortName`, and `Type`. We then derive a set of predicates, each of which expresses the semantics of one of the views using the 'language' defined earlier. As an example, List. 3 shows the predicates for the C&C views `RJFunction` and `RJStructure` taken from specification $S_1$ (Fig. 1).

Finally, we construct a predicate `spec` representing the specification's Boolean expression over the C&C views' predicates.

### 4.3 Synthesis

We run the module defined above with a command that tries to satisfy the `spec` predicate. Note that Alloy analysis must be done within a user-defined, given scope, which specifies an upper bound for the number of instances per signature. In our case, the upper bound for the number of components is derived from the specification. The upper bound for the total number of ports and the maximal number of ports per component, however, cannot be derived from the specification. We thus let the user choose the scope for ports (we can only derive a lower bound for the number of necessary ports).

If an Alloy instance is found, we translate it back to the problem domain, that is, to a complete C&C model. The translation back is straightforward and is linear in the size of the solution. However, if an Alloy instance is not found, in the general case, we do not know whether the specification could be satisfiable in a larger scope, that is, using more ports. As stated before, our solution is indeed sound but incomplete (although it is complete within the given scopes).

## 5. ADVANCED FEATURES

### 5.1 Library Components

Most C&C models reuse library components, pre-defined or existing components adopted from other systems. Thus,

it is crucial that our technique would allow the engineer to import such components and apply an integrated synthesis solution.

Specifically, C&C views synthesis supports the integration of library components or similar components at two levels. First, the engineer can extend the specification with a list of imported library component definitions. A component definition is complete: it specifies the complete interface (port names and types) of the component. If a component definition for component `Cmp` is imported, we check that none of the C&C views mentioning `Cmp` explicitly specifies a subcomponent for `Cmp`, and we add its interface (and the requirement that it is complete) as an additional constraint to the generated Alloy module. This ensures that a synthesized design that uses `Cmp`, if any, would be consistent with its interface and use it as is.

As an example, consider the `ServoValve` component in specification $S_1$ from Fig. 1 to be an imported library component. In the library, the complete interface of `ServoValve` is given. Importing it to the specification $S_1$ ensures its interface is used as is in the synthesized design and rules out solutions that put other components that are mentioned in the C&C views as its subcomponents. For example, a solution where `ServoValve` contains `Sensor` will not be possible.

The additional Alloy predicates we use to support library components are shown in List. 4 and List. 5. We instantiate these predicates with every library component in the specification, as shown in List. 6 for the `ServoValve`.

Note that the synthesis considers an imported component as a black-box: it uses its interface and needs no knowledge of its subcomponents. This is meant to support encapsula-

```
1  pred RJFunction {
2    one RotationalJoint
3    and one Actuator
4    and one Sensor
5    and one Cylinder
6    and independentSet[Sensor + Actuator + Cylinder]
7    and contains[RotationalJoint, Sensor + Actuator
8                                  + Cylinder]
9    and connected[Sensor, Actuator] }
10
11 pred RJStructure {
12   one RotationalJoint
13   and one Body
14   and one ServoValve
15   and one Cylinder
16   and independentSet[ServoValve + Body + Cylinder]
17   and contains[RotationalJoint, ServoValve + Body
18                                  + Cylinder]
19   and portOfComponent[Cylinder, IN, my_float,
20                                              angle]
21   and connected[RotationalJoint, ServoValve]
22   and connected[ServoValve, Body]
23   and connectedWithReceiverPortName[Body,
24                                    Cylinder, angle]
25   and connected[Cylinder, Body] }
```

**Listing 3:** Predicates for two C&C views, `RJFunction` and `RJStructure`, from specification $S_1$ (see Sect. 4.2)

```
1 pred libraryComponent[cmp: Component] {
2   no cmp.subComponents }
```

**Listing 4:** Predicate to specify that a library component must have no subcomponents (see Sect. 5.1)

tion and modularity: as long as the interface is kept fixed, the implementation of imported library components can be replaced without affecting the synthesized design.

Second, the designer can strengthen a C&C view by declaring some of the components mentioned in it as *interface complete* (technically, using a stereotype). This is useful when the designer knows the complete interface of a component she is using although this component is not a black-box library component. For example, consider adding the stereotype interface complete to the `Sensor` in the `SensorConnections` model (Fig. 1). This would mean that in the synthesized design, `Sensor` must have exactly the set of ports and corresponding types shown in this model. As `Sensor` is not imported as a library component, it is still possible that in the synthesized solution it will include subcomponents from the set of components already mentioned in the models.

The additional Alloy predicate we use to support interface complete components is shown in List. 5. We instantiate this predicate with every component that is specified as interface complete in a model in the specification.

### 5.2 Specification Patterns

The use of a Boolean expression in C&C views specifications makes them very expressive. However, for the engineer who constructs the specification, using only low level basic Boolean connectives may be inconvenient and error prone. Thus, we look for higher-level specification patterns, which can be used to express the required semantics more intuitively, and can be reused across different specifications.

Based on our experience with creating specifications, some examples of simple patterns are: [ALT], given a set of C&C views, specifying that the synthesized design must satisfy at least one of the models in the set; [XALT], given a set of views, specifying that the synthesized design must satisfy

```
1 pred interfaceComplete[cmp:Component,
2         portNames:set PortName] {
3   cmp.ports.name = portNames }
```

**Listing 5:** Predicate to specify that a component's interface is complete, technically, by stating that the set of its port names, as appearing in the model, is exactly its complete set of port names (see Sect. 5.1)

```
1 fact libraryComponents {
2   libraryComponent[ServoValve]
3   some ServoValve implies (
4     portOf[ServoValve, IN, my_float, portIn]
5     and portOf[ServoValve, OUT, my_float, portOut]
6     and interfaceComplete[ServoValve,
7       portIn + portOut]) }
```

**Listing 6:** A fact stating that `servoValve` is a library component (see Sect. 5.1). If `servoValve` is used in the design its interface is completely specified.

exactly one of the views in the set; [IMP], given two views, if the design satisfies the first, it should also satisfy the second; and [NOCOMP], a given component should not be present in the synthesized design. Note that the last three patterns depend on the use of negation in the language.

As an example, recall specification $S_1$ (Fig. 1). Assume that the engineer responsible for `Sensor` knows that if it is located outside `Body`, then it must use an inner amplifier `Amplifier`. To express this knowledge, the architect can create the C&C view `SensorHasAmp` shown in Fig. 4 (left) and add the implication IMP(BodySensorOut,SensorHasAmp) to the specification's Boolean expression (as another conjunct). To make sure the amplifier is not used when it is not necessary, the engineer can create the view `Amp` shown in Fig. 4 (right), and add the implication IMP(BodySensorIn,¬Amp) (since the model `Amp` consists of the single component `Amplifier`, the expression ¬Amp used here is an instance of the pattern NOCOMP mentioned above).

As another example, the choice between `BodySensorIn` and `BodySensorOut` in $S_1$ can be strengthened by the engineer to XALT(BodySensorIn,BodySensorOut) without loosing or adding possible implementations, that is, as a form of refactoring. The semantics of the two views entails that no design can satisfy both and their use in the specification $S_1$ entails that at least one of them must be satisfied.

Specification patterns do not add expressive power. Rather, they are only meant to improve the readability and usability of C&C views specifications. Indeed, as part of our prototype implementation, we have developed graphical interfaces to add and edit the specification's Boolean expression, including support for several patterns. Our ideas on specification patterns for architecture structures are inspired by previous works on patterns of temporal specifications (e.g., [8]).

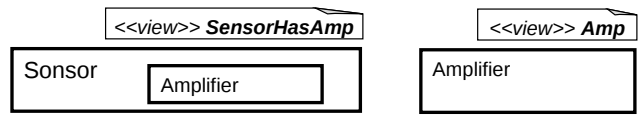

**Figure 4:** The additional C&C views `SensorHasAmp` and `Amp`, to be added to the specification $S_1$ with the implications IMP(BodySensorOut,SensorHasAmp) and IMP(BodySensorIn,¬Amp) (see Sect. 5.2).

```
1  fact hierarchicalArchitecture {
2    no c: Component |
3      c in ^(Talker.talksTo).c }
4
5  one sig Talker {
6    talksTo : Component -> Component
7  } { all c1, c2: Component |
8      c2 in talksTo.c1 iff endToEndConnection[c1,c2]}
9
10 pred endToEndConnection[senderC : one Component,
11                         receiverC: one Component] {
12   some senderP : senderC.ports |
13            some receiverP : receiverC.ports |
14     no senderP.sendingPort and // no forward
15     no receiverP.receivingPorts and // no forward
16     receiverP in senderP.^receivingPorts }
```

**Listing 7:** Excerpts from the Alloy code for the hierarchical style (see Sect. 5.3)

## 5.3 C&C Views Synthesis with Styles

Architectural styles systematize successful architectural design practices in terms of constraints on architectural elements and their composition into systems [29]. Some examples of well-known architectural styles that are relevant to C&C models include the pipe-and-filter style, the client-server style, and the layered style (an overview can be found in Chap.4 of [30]). We have extended C&C views synthesis with support for three architectural styles. As applicable, style specific constraints are added to the synthesis input so that the synthesized design, if any is found, obeys the rules of the style. We give three examples below.

First, a *hierarchical style*, whose essence is to forbid the C&C models from having directed cycles of connected components. This style is important, because hierarchical architectures are suitable for behavioral synthesis: while the problem of synthesizing a finite-state distributed reactive system over a given architecture is in general undecidable, it is decidable for the class of hierarchical architectures [28].

To enforce the hierarchical style of the synthesized model we add to the generated Alloy module a fact that requires that no directed cycles exist (see List. 7). More technically, this is implemented using the helper relation `talksTo` of the signature `Talker`. A pair of components $C_1, C_2$ is in this relation iff there exists a connection that starts at $C_1$ (not only forwarded) and ends at $C_2$. If a hierarchical solution exists, it will be found. If no solution exists, it might be possible to find a solution that is not hierarchical.

Second, a *client-server style*, whose essence is to identify one of the components as a single server and to forbid any direct communication between clients. Also, the server and the clients are assumed all to be independent in terms of containment; a client is not contained within the server or another client etc. To integrate this style with our synthesis solution we add the identities of the server and the client components, as defined by the engineer, as additional input to the specification. We replace the constraint of a single top component (see Sect. 3.1), with a constraint that specifies that the server and the clients are the top components. We then use a fact to enforce additional style constraints on the synthesized design (see List. 8). More technically, the fact uses a parametrized predicate that defines when two components are directly connected, and two generated functions, `myServer` and `myClients`, returning the server component and the client components resp. Again, if a design that satisfies the specification together with the additional restrictions exists, it is found. If not, it means that the semantics of the specification cannot be satisfied within the client-server style, e.g., without direct client to client connectors (or that it cannot be satisfied at all).

Third, a *layered style*, which forces a partition of the components into a sequence of layers, and allows direct connectors only within layers and between consecutive layers. To integrate this style with our synthesis solution we add the partition of the components into layers, as defined by the engineer, as additional input to the specification. As in the client-server style, we replace the constraint of a single top component with a constraint that specifies that the layers are the top components. We then use several parametrized predicates to enforce the style's constraints on the synthesized design. If a design that satisfies the specification together with the additional restrictions induced by the partition of the layers exists, it is found. If not, it means that the partition into layers conflicts with the semantics of the specification (or that the specification, even without the style constraints, is not satisfiable, at least within the given scope).

It is important to note that each of the C&C views in the specification is independent of and does not have to be compliant, by itself, with the constraints induced by the architectural style. For example, even though a layered architecture is enforced, abstract connectors in a view may connect components from nonconsecutive layers. The synthesis is responsible to implement these abstract connectors through chains of concrete connectors that obey the layered architecture. As another example, what looks like a communication cycle in a given view (and thus apparently violates a hierarchical style), may end up implemented in the synthesized design without creating a cycle. For example, recall specification $S_1$ of Fig. 1 where view `Structure` seems to contain a cycle between `Body` and `Cylinder`. In the synthesized design, shown in Fig. 2, we see that the implementation contains no concrete end-to-end cycle. Thus, the C&C views and the architectural style are specified independently. The synthesis is responsible for finding a design that satisfies both, if one exists.

Finally, the use of architectural styles adds expressive power to C&C views specifications and makes it applicable to a wide range of systems. In our prototype plug-in for C&C views synthesis, we have added graphical user interfaces to select a style and edit its properties, including support for the three styles described above.

```
1  fact clientServerArchitecture {
2    all client : myClients |
3      immediatelyConnectedNoOrder[myServer, client]
4      and no c : (myClients-client) |
5      immediatelyConnectedNoOrder[client, c] }
6
7  pred immediatelyConnectedNoOrder[c1: Component,
8                                   c2: Component] {
9    some p : c1.ports |
10     ( p in c2.ports.receivingPorts or
11       p in c2.ports.sendingPort ) }
```

**Listing 8:** Excerpts from the Alloy code for the client-server style. `myServer` and `myClients` are generated Alloy functions returning the server component and the client components resp. (see Sect. 5.3)

## 6. IMPLEMENTATION AND EVALUATION

The plug-in implementation and all specifications reported on below are available in supporting materials [3], together with screen captures and relevant documentation. All specifications can be inspected and all experiments can be reproduced. We encourage the interested reader to try it out.[2]

### 6.1 Implementation

We implemented C&C views synthesis in an Eclipse plug-in prototype. The input consists of a C&C views specification. The specification — selection of C&C views and components, definition of the Boolean formula, scope, and optional parameters about styles — is edited using a dedicated graphical UI. The concrete syntax used is adapted from MontiArc [2]. At the back end, the plug-in implements the translation to Alloy using MontiCore APIs [23] and FreeMarker [10]. The SAT solver we use is MiniSat [27]. When a design is synthesized, it is presented as a MontiArc document to the engineer, who can inspect it and further use it for code generation (to Java or to MathWorks Simulink).

We tested the implementation over C&C views specifications for four systems from different sources and of different domains. We experimented with several different specifications for each system, in order to test and evaluate the use of different features (e.g., library components, specification patterns, architectural styles). For validation, we have also implemented a polynomial algorithm that checks whether a given C&C model satisfies a given C&C view. As part of our experiments, we ran this algorithm on the results of all successful synthesized designs and their source views. This check is available with the prototype plug-in too.

### 6.2 Evaluation

We evaluated C&C views synthesis on four systems, taken from different sources.

**Avionics system.** We evaluated C&C views synthesis on an AADL architecture of an avionics system, taken from [4] (specifically `Avionics_System.aadl` of the OSATE AADL Project). The avionics system architecture is a high-level model of several avionics system subsystems.

Based on various use cases related to interactions between system's components, we created 9 C&C views, 1-6 components each. For example, one view gives an overview of the complete data flow in the system, declared using abstract connectors. This view does not provide additional information such as port names or types. Another view provides more details about the communication between the `Pilot_Display` and it's `Page_Content_Manager`, showing incoming and outgoing ports with their names and connectors. We defined 7 satisfiable and unsatisfiable C&C views specifications, using 3 to 9 views, some extended with styles.

**Pump station.** We further experimented with a pump station design taken from an example system provided with the AutoFocus tool, developed at TU Munich [1, 18]. The physical pump station system consists of two water tanks connected by a pipeline system with a valve and a pump. The water level in the first water tank can rise (this is controlled by the environment). When the water level of the first tank rises to a critical level, the water has to be pumped to the the second water tank. The second water tank has a drain.

Based on several design decisions and relations we wanted to highlight and document, we created 10 C&C views, each with 2-5 components. For example, one view gives an overview of the basic structure of the system and omits details about interfaces and connectors. Another view documents part of the connections between the actuators and their environment, hiding hierarchies and omitting elements not connected to the actuators. An additional C&C view shows an undesired design where the simulation component is placed inside the pumping system.

We defined 8 satisfiable and unsatisfiable C&C views specifications. Two specifications specify the optional existence of an emergency system and its implications, using the [ALT] and [IMP] patterns. Another specification prohibits an emergency system. Other specifications combine models of the function of the pump station with ones that specify the separation of the pumping system from the simulation part.

**Robotic arm.** We evaluated C&C views synthesis on a robotic arm architecture – specifically a rotational joint, taken from an industrial system by VTT Tampere, Finland (the system used as running example in this paper). The main components of the rotational joint's model are a cylinder, a servo valve, a sensor, a joint limiter, and an actuator. The rotational joint is a subsystem of a robotic arm containing 8 rotational (identical copies) and translational joints.

Based on several requirements and partial knowledge or particular features, we created 11 C&C views, each with 1-5 components. Some views highlight the components necessary for the function of the joint while others document design alternatives on the placement of sensor and actuator components. Some of the views give an overview over related components with only few details of their interfaces or connectedness. Other views document complete interfaces of relevant components and some of their connections. Moreover, we created 8 C&C views specifications, each combining 6-8 C&C views to express design alternatives (pattern [ALT]), undesired designs, and implications of design decisions (pattern [IMP]).

**Lunar lander.** We evaluated C&C views synthesis on the lunar lander model, which is used by Taylor et al. as a running example in their book on software architecture [30] and presented in a related work by Bagheri and Sullivan [6]. The lunar lander is space ship with various sensors, a controller, and actuators. The objective of the lunar lander is to land safely on the surface of the moon.

Based on the natural language description of the lunar lander consisting of three components presented in [30, pp. 201] we have created 8 C&C views, each with 1-3 components. Each C&C view covers parts of the natural language description. Since this description is formulated positively, the C&C views specification is a conjunction of the 8 views.

Based on a shorter natural language description of the same lunar lander with 8 components (4 sensors, one controller, and 3 actuators) from [6] we have created 2 views and again the C&C views specification is a conjunction of these 2 views. Both original examples from [30] and [6] contained neither port names nor types, and no hierarchy. We introduced port names in the C&C views of the 3 component lunar lander example.

### 6.3 Observations and Lessons Learned

**Performance.** For some specifications, synthesis took only a few seconds, while for others it took up to a minute to complete (on a regular laptop computer). Our experience shows that relatively minor changes in a specification, such as ones that add no C&C views or components but only further constrain the specification with library components or statements of interface completeness, and even ones that do not affect the semantics (such as different ordering of the models in the Boolean formula), sometimes have a significant effect on performance. Indeed, it is known that Alloy's performance, and SAT solvers' performance in general, are sensitive to the order of variables in their input.

---

[2]The supporting materials have been successfully evaluated by the ESEC/FSE artifact evaluation committee and found to meet expectations.

Since C&C views synthesis is NP-hard, one cannot expect it to be instantaneous. Thus, we consider the resulting times to be reasonable. As we have not made special efforts to optimize the implementation, we believe there is much room for improvement in this regard. Also, interestingly, deciding that a specification is not satisfiable was typically (but not always) much faster than deciding that it is satisfiable and providing the synthesized design.

**Language expressiveness.** We found that the use of a Boolean formula over C&C views, together with the patterns, in particular the use of alternatives, negations, and implications, is both expressive and easy to read and write. We thus believe that C&C views' 'by example' characteristics is attractive to engineers (our belief is supported also by the analogous use of scenarios in behavioral specifications). On the other hand, the use of C&C views to specify some properties was not always natural and intuitive. Instead, sometimes we wished to have a more fine-grained, flexible, and powerful language that allows one to write symbolic, succinct specifications (e.g., using quantification). We leave this topic for future work.

**Multiple solutions.** A C&C views specification may have more than one satisfying design. For example, consider the specification $S_1$ of Fig. 1. Fig. 2 shows a possible solution. An alternative satisfying design may be an architecture which is identical to the first except that `ServoValve` contains `Sensor`. If this is not an acceptable design, the architect can disallow it (e.g., by adding it, or a smaller C&C views consisting of the `ServoValve` and the `Sensor`, in a negated form to the specification), and run synthesis again.

Most of our specifications had many satisfying solutions. So, we found that it may be useful to have a better way of choosing between solutions, e.g., by optimizing some cost functions (hierarchy depth, ports' number, connector chains length). This however cannot be efficiently done using our current technique. We leave this issue for future work.

**Handling unsatisfiable specifications.** A C&C views specification may be unsatisfiable. In addition to identifying unsatisfiability, one may be interested in presenting the root cause of conflicts to the engineer, i.e., to identify a minimal subset of the specification (Boolean formula) that is unsatisfiable. Note that unsatisfiability may have more than one cause, for example, as explained earlier in Sect. 2.2, the specification $S_2$ is unsatisfiable both because of a conflict in component's containment and a port type mismatch.

Heuristics may be used to detect some simple patterns of unsatisfiability in linear or polynomial time. A complete but inefficient solution to find a minimal subset would require an exponential number of synthesis runs. We believe that providing tools to handling unsatisfiable specifications efficiently is an important direction for future work. One may be able to build these on top of existing technologies for UNSAT core, as supported by some SAT solvers. However, in order to be effective, the identified core must be lifted and presented to the engineer back using the abstractions defined by the C&C views.

## 7. RELATED WORK

We discuss related work in the areas of component and connector modeling and analysis and in the area of behavioral specifications. The unique features of our work on C&C views and synthesis are its support for hierarchy, its support for abstraction of direct hierarchy, direct connectivity, names and types, its focus on structure, and its goal, to automatically construct a satisfying C&C model from a given set of constraints which are expressed as views and crosscut the implementation oriented decomposition of systems to sub-systems.

Some previous works deal with the analysis or synthesis of component and connector structures, mainly in the context of architectures [6, 7, 13, 19, 21]. We discuss these works and compare them to our work below.

Bagheri and Sullivan [6] present an approach to synthesize style specific architecture descriptions from a style independent application description. Our work includes support for architecture styles too (see Sect. 5.3). Unlike our work, the input of their Alloy-based synthesis is a *single and complete application description* and not a specification over multiple, incomplete and crosscutting models like C&C views. In [6], the application description is mapped to multiple possible architecture descriptions in a specific style. It is unclear whether the mapping (synthesis) can fail (besides from choosing too small an Alloy scope). Finally, unlike our support for hierarchical structures, the application description and architecture styles presented in [6] do not support hierarchy.

Bagheri and Sullivan [6] use the lunar lander example as a case study. Although this example is small and flat (and thus does not take advantage of the unique features of our approach), we use it in our evaluation too, for the purpose of comparison (see Sect. 6).

Kim and Garlan [21] present formal analysis of C&C models within architectural styles, using Alloy. Unlike our work, however, their work does not support hierarchical decomposition of components (their configurations are flat). One analysis supported is "checking the constructability of specific architectural configurations" (Sect. 7.2): the input is a (partial) architecture configuration and the output is an instance of an architecture in the given style that contains the input configuration. The input in [21] consists of a single configuration and not a set of separated views as supported in our work. Moreover, [21] does not support abstraction of hierarchy or connectivity as is available in our work.

Issarny et al. [19, 22] present a development environment for the composition of middleware architectures. The input for composition are two C&C models. The output of the directed composition operator is a single C&C model where possibly multiple copies of components of the second model are connected with components of the first. A valid composition preserves communication order of the input models and CTL properties checked against PROMELA implementations of components, both specified by the user. The work may be viewed as similar to ours, as it translates a C&C model composition problem into a model checking problem. It differs significantly, since the input in [19, 22] consists of concrete models without common elements, with-out hierarchies (they are flat) or abstraction mechanisms as supported by C&C views. Moreover, their composition is binary and directed while ours supports Boolean specifications over C&C views, including negation, alternatives etc.

Boucké et al. [7] present composition operators for C&C models to avoid the repetition of elements in integrated models. The integration of sets of architecture models is based on a user specified unification relation (marking identical elements across input models) and a submodel relation (e.g., a component is detailed in another model). This model composition is thus an imperative composition of concrete C&C models rather than a declarative one as in our approach. We believe that our work may benefit as well from a more complex and maybe user specified unification relation across C&C views as suggested in [7].

Giese and Vilbig [13] discuss separation of non-orthogonal concerns in software architecture and design. The work deals with composition of structure as well as of behavior from architectural views. Architectural views are defined as directed graphs representing components and connectors extended with behavior contracts. According to [13], the structural part of their architectural view synthesis can be handled by superposition of these directed graphs. In contrast, the synthesis problem for our C&C views specifications is more complex because of its rich abstraction mechanisms (of direct hierarchy and connectivity) and its support for the specification of negative or alternative designs, which prohibit simple superposition.

AADL [9] includes under-specification mechanisms similar to the ones available in our work. For example, AADL supports specifications with incomplete information of port types and with abstract flows, which show the source and sink of flows but not their complete path through the system. In [9], Feiler et al. explain that the motivation behind AADL's support for partial specifications is to allow some analysis (e.g., computing some metrics, checking syntactic correctness) already during early design, before all implementation details are known. However, a synthesis of a complete architecture from a set of partial specifications, as is done in our work, is not discussed.

Acme [12] allows one to specify first-order predicates on the structure of architectures, dealing, e.g., with connectedness between components. In AcmeStudio [5], these predicates are automatically evaluated against the architecture. The predicates are written as logical formulas, not as abstract models such as C&C views. Moreover, we have not seen any work that suggests to use these predicates as input for synthesis of a satisfying architecture.

AspectualAcme [11] extends Acme with the modular representation of architectural aspects and their multiple composition forms. The extension allows special aspectual components to be connected to other components via special aspectual connectors, characterized by aspectual glues of before, after, and around types (with a semantics similar to that of advice composition in AspectJ). This work and ours are remotely related, specifically in the idea of crossing the hierarchical boundaries of components and subcomponents. However, the works are also very different, as AspectualAcme deals with the extension of a given base architecture and its fixed structure with crosscutting concerns, while our work focuses on using declarative views as specification for the structure of the model and on the use of these views for its synthesis.

Previous work in our group by Gronniger et al. [14, 15] described the use of views in the context of product lines, with a focus on the automotive domain, using SysML internal block diagrams. Like AADL, these provide underspecification mechanisms, e.g., to specify abstract connectors. Synthesis from views was not discussed in these works.

Finally, from a broader, higher-level perspective, variants of synthesis from partial views have been extensively studied in the area of behavioral specifications (see, e.g., the works of Harel and Segall., Maoz and Sa'ar, Uchitel et al., and Whittle and Schumann on synthesizing behavior models from scenarios [17, 24, 31, 32]). Most importantly, however, these works assume the structure of the systems to be given; only the behavior is synthesized!

Our work on C&C views deals with the synthesis of structures, not behaviors, and may be viewed as complementing some of these works. As future work one may develop integrated structural and behavioral synthesis techniques.

## 8. CONCLUSION

We presented component and connector views, which specify structural properties of component and connector models in an expressive and intuitive way. C&C views provide means to abstract away direct hierarchy, direct connectivity, port names and types, and thus can crosscut the traditional boundaries of the implementation-oriented hierarchical decomposition of systems and sub-systems, and reflect the partial knowledge available to different stakeholders involved in a system's design.

As a primary application for C&C views we investigated the synthesis problem: given a C&C views specification, consisting of mandatory, alternative, and negative views, construct a concrete satisfying C&C model, if one exists. We showed that the problem is NP-hard and solved it, in a bounded scope, using a reduction to SAT, via Alloy. We further extended the basic problem with support for library components, specification patterns, and architectural styles. The result of synthesis can be used for further exploration, simulation, and refinement of the C&C model or, as the complete, final model itself, for direct code generation.

Several future research directions arise from our evaluation in Sect. 6, including strengthening the expressiveness of C&C views with parametrized component instantiation or quantification, and better handling the case where the specification is unsatisfiable by providing directions to the root cause of unsatisfiablity (as is done in the behavioral case when a set of scenarios is unrealizable, see [25]).

We believe that all the above can be defined and implemented on top of the C&C views language and the synthesis solution presented in our work. Our ultimate future work goal is to integrate structural and behavioral synthesis into a single, comprehensive solution for C&C models specification and implementation.

## 9. ACKNOWLEDGMENTS

We thank S. Szekely and G. Weiss for advice on Eclipse plug-in development. We thank A. Haber for discussions on C&C views and help with MontiArc API. J.O. Ringert is supported by the DFG GK/1298 AlgoSyn.

# 10. REFERENCES

[1] AutoFocus3 website.
http://autofocus.informatik.tu-muenchen.de/.
Accessed 2/2013.

[2] MontiArc website.
http://www.monticore.de/languages/montiarc/.

[3] Supporting materials on C&C views synthesis.
http://www.se-rwth.de/materials/cncviews/.

[4] AADL website. http://www.aadl.info/. Accessed 2/2013.

[5] The ACME Studio Homepage.
http://www.cs.cmu.edu/~acme/AcmeStudio/.
Accessed 2/2013.

[6] H. Bagheri and K. J. Sullivan. Monarch: Model-based development of software architectures. In *MoDELS (2)*, volume 6395 of *LNCS*, pages 376–390. Springer, 2010.

[7] N. Boucké, D. Weyns, and T. Holvoet. Composition of architectural models: Empirical analysis and language support. *Journal of Systems and Software*, 83(11):2108–2127, 2010.

[8] M. B. Dwyer, G. S. Avrunin, and J. C. Corbett. Patterns in property specifications for finite-state verification. In *ICSE*, pages 411–420, 1999.

[9] P. H. Feiler, D. P. Gluch, and J. J. Hudak. The architecture analysis & design language (AADL): An introduction. Technical report, Software Engineering Institute, Carnegie Mellon University, 2006.

[10] FreeMarker website.
http://freemarker.sourceforge.net/. Accessed 2/2013.

[11] A. Garcia, C. Chavez, T. V. Batista, C. Sant'Anna, U. Kulesza, A. Rashid, and C. J. P. de Lucena. On the modular representation of architectural aspects. In V. Gruhn and F. Oquendo, editors, *EWSA*, volume 4344 of *Lecture Notes in Computer Science*, pages 82–97. Springer, 2006.

[12] D. Garlan, R. T. Monroe, and D. Wile. Acme: Architectural description of component-based systems. In *Foundations of Component-Based Systems*, pages 47–68. Cambridge University Press, 2000.

[13] H. Giese and A. Vilbig. Separation of non-orthogonal concerns in software architecture and design. *Software and Systems Modeling*, 5(2):136–169, 2006.

[14] H. Grönniger, J. Hartmann, H. Krahn, S. Kriebel, L. Rothhardt, and B. Rumpe. Modelling automotive function nets with views for features, variants, and modes. In *ERTS*, 2008.

[15] H. Grönniger, J. Hartmann, H. Krahn, S. Kriebel, L. Rothhardt, and B. Rumpe. View-centric modeling of automotive logical architectures. In *MBEES*, volume 2008-2 of *Informatik-Bericht*, pages 3–12. TU Braunschweig, Institut für Software Systems Engineering, 2008.

[16] A. Haber, J. O. Ringert, and B. Rumpe. Montiarc - architectural modeling of interactive distributed and cyber-physical systems. Technical Report AIB-2012-03, RWTH Aachen, February 2012.

[17] D. Harel and I. Segall. Synthesis from scenario-based specifications. *J. Comput. Syst. Sci.*, 78(3):970–980, 2012.

[18] F. Hölzl and M. Feilkas. Autofocus 3 - a scientific tool prototype for model-based development of component-based, reactive, distributed systems. In *Model-Based Engineering of Embedded Real-Time Systems*, volume 6100 of *LNCS*, pages 317–322. Springer, 2007.

[19] V. Issarny, C. Kloukinas, and A. Zarras. Systematic aid for developing middleware architectures. *Commun. ACM*, 45(6):53–58, 2002.

[20] D. Jackson. Alloy: a lightweight object modelling notation. *ACM Trans. Softw. Eng. Methodol.*, 11(2):256–290, 2002.

[21] J. S. Kim and D. Garlan. Analyzing architectural styles. *Journal of Systems and Software*, 83(7):1216–1235, 2010.

[22] C. Kloukinas and V. Issarny. SPIN-ning Software Architectures: A Method for Exploring Complex. In *WICSA*, pages 67–76. IEEE Computer Society, 2001.

[23] H. Krahn, B. Rumpe, and S. Völkel. MontiCore: a framework for compositional development of domain specific languages. *STTT*, 12(5):353–372, 2010.

[24] S. Maoz and Y. Sa'ar. Assume-guarantee scenarios: Semantics and synthesis. In R. B. France, J. Kazmeier, R. Breu, and C. Atkinson, editors, *MoDELS*, volume 7590 of *Lecture Notes in Computer Science*, pages 335–351. Springer, 2012.

[25] S. Maoz and Y. Sa'ar. Counter play-out: executing unrealizable scenario-based specifications. In *ICSE*, pages 242–251. IEEE / ACM, 2013.

[26] MathWorks Simulink website.
http://www.mathworks.com/products/simulink/.
Accessed 2/2013.

[27] MiniSat website. http://minisat.se/. Accessed 2/2013.

[28] A. Pnueli and R. Rosner. Distributed reactive systems are hard to synthesize. In *FOCS*, pages 746–757. IEEE Computer Society, 1990.

[29] M. Shaw and D. Garlan. *Software architecture - perspectives on an emerging discipline*. Prentice Hall, 1996.

[30] R. N. Taylor, N. Medvidovic, and E. Dashofy. *Software Architecture: Foundations, Theory, and Practice*. Wiley, 2009.

[31] S. Uchitel, G. Brunet, and M. Chechik. Behaviour model synthesis from properties and scenarios. In *ICSE*, pages 34–43, 2007.

[32] J. Whittle and J. Schumann. Generating statechart designs from scenarios. In *ICSE*, pages 314–323. ACM, 2000.